\documentclass[11pt,a4wide]{article}
\usepackage{newlfont}
\usepackage{amsmath}
\usepackage{graphics}
\usepackage{float}
\usepackage{graphicx}
\usepackage{epsfig}

\begin{document}
\title{Higher Hybrid Charmonia in an Extended Potential Model}
\author{M. Atif Sultan \thanks{e mail: atifsultan.pu@gmail.com}
\quad, Nosheen Akbar\thanks{e mail: nosheenakbar@ciitlahore.edu.pk}\quad,
Bilal Masud\thanks{e mail: bilalmasud.chep@pu.edu.pk}\quad, Faisal Akram \thanks{e mail: faisal.chep@pu.edu.pk}\\
\textit{Centre For High Energy Physics, University of the Punjab,
Lahore(54590), Pakistan.}\\
\textit{$\dag$COMSATS Institute of Information and Technology,
Lahore(54000), Pakistan.}}
\date{}
\maketitle

\begin{abstract}

   The quark potential model for mesons and its extension for hybrid mesons are used to study the effects of radial excitations on the masses, sizes and radial wave functions at the origin for conventional and hybrid charmonium mesons. These results can help in experimentally recognizing  hybrid mesons.
    The properties of conventional and hybrid charmonium mesons are calculated for the ground and radially excited states using the shooting method to numerically solve the required Schr$\ddot{\textrm{o}}$dinger equation for the radial wave functions. We compare our results with the experimentally observed masses and theoretically predicted results of the other models. Our results have implications for scalar form factors, energy shifts, and polarizabilities of conventional and hybrid mesons.
    The comparison of masses of conventional and hybrid charmonium meson with the masses of recent discovered XYZ-particles is also discussed.
\end{abstract}

\maketitle
\section*{I. Introduction}

In conditions where the established theory of hadronic physics, namely quantum chromodynamics (QCD), cannot be solved we use a variety of models to explain the hadronic properties. To test these models, we can compare them with numerical simulations of QCD, like those through the lattice gauge theory, and
available results of hard experiments. (The effort remains to use these discrete tests of continuum models for improvements in the models so as to get better agreements in the next comparisons. This is an important route of advancing our understanding of the hadronic physics and QCD.) Such hard experiments could be cross sections, decay rates, masses and $J^{PC}$ (angular momentum, parity and C-parity) combinations of hadrons. In this paper we present a comprehensive list of phenomenological implications in the charmonium sector, in form of masses, radii and wave functions at origin for a variety of $J^{PC}$ assignments, of a model previously proposed \cite{Nosheen11} by us for hybrid mesons. This model is an extension of the quark potential model to incorporate the knowledge of the gluonic excitations provided by lattice gauge theory; this analytic model was noted to have a very good agreement with the corresponding lattice gauge theory based numerical results. If this QCD-motivated model can explain properties of newly discovered mesons including many hybrid candidates, this should be a useful advancement in our understanding of the gluonic excitations  and, in general, of the physics beyond in the quark model. In our quest for comprehensiveness, we address the radial and orbital
excitations along with the previously worked out \cite{Nosheen11} radially ground state gluonic excitations (hybrids) and try to explain the properties of the newly discovered mesons in the charmonium sector mentioned below.

In general, an important guide to the search of physics beyond the quark model are meson $J^{PC}$ combinations like
$0^{--},0^{+-},1^{-+},2^{+-}$, and $3^{-+}$ \cite{Barnestalk} not satisfying the quark model formulas $P =(-1)^{L+1}$ and $C = (-1)^{L+S}$. Amongst the theoretical options to understand such exotic mesons, hybrid is an important option with often some expected features predicted through the existing models of hybrids or numerical simulations of the excited gluonic field in QCD. Guided by theoretical signatures for hybrids, it is not uncommon to find experimentalists specifically searching for hybrid mesons:  CERN COMPASS has been \cite{Moinester00} centered on hybrid meson structure studies, along with pion polarizability. Progress on this project is reported, for example, in \cite{COMPASS later}. Hybrid interpretations of some results form $B$-factories has been discussed in \cite{Belle}. Ref. \cite{Carman06} discussed then available experimental evidence for exotic hybrid mesons and described the GlueX experiment in Hall D of Jefferson laboratory as a new initiative for performing detailed spectroscopy of the light-quark meson spectrum. This collaboration aims to investigate the full spectrum of mesonic states upto roughly 3 GeV including hybrid candidates; the photoproduction in it promises to be rich in hybrids, starting with those having exotic quantum numbers. Hence this experiment has been expected to provide detailed spectroscopy necessary to map out hybrid meson spectrum
which is essential for an understanding of the confinement mechanism and the nature of the gluon in QCD. This primary motivation of GlueX has been re-stated in more recent proposals \cite{GlueXproposals} as to search and ultimately study the pattern of gluonic excitations in the meson spectrum produced in $\gamma p$ collisions. The spectroscopic searches for the exotic hybrid states such as $c g\bar{c}$ system in the CLEO-c Research Program \cite{Cleo}. The PANDA (antiProton ANnnihilation at DArmstadt) experiment features a scientific programme devoted \cite{PANDA} to charmonium spectroscopy and gluonic excitations (hybrids, glueballs), along with some other topics. This experiment performs studies of the strong interaction via precision spectroscopy
of charmonium and open-charm states, an extensive search for exotic objects such as
glueballs and hybrids, in-medium and hypernuclei spectroscopy, and more. Ref. \cite{arxivbib} says ``The research of charmonium and charmed hybrids
using the antiproton beam with momentum ranging from 1 GeV/c to
15 GeV/c in PANDA experiment at FAIR is perspective and interesting from the scientific
point of view. Charmonium and charmed hybrids with different quantum numbers are
copiously produced in antiproton-proton annihilation process." Ref. \cite{Wiedner} says that for the PANDA experiment ``Its set-up allows physicists to address questions like the structure of glueballs and hybrids, to clarify
the nature of the X, Y, and Z states" and other related ones. This also tells that Crystal Barrel searched for hybrids with the exotic quantum numbers $J^{PC} = 1^{-+}$.

Many exotic mesons are suggested to be hybrids. For example, ref. \cite{compas}  mentions three experimental candidates for a light $1^{-+}$ hybrid:  $\pi_1(1400)$, observed by E852, VES, and Crystal Barrel experiment, $\pi_1(1600)$, observed by E852 and VES, and $\pi_1(2000)$ observed by E852; see also the
ref. \cite{Lu2005} for the latter two mesons, which also describes $\pi_2(1880)$ as a hybrid meson candidate. Ref. \cite{Ding Yan} proposes that a
structure at 2175 MeV observed by the Babar collaboration is a $1^{--}$ strangeonium hybrid. The first paper of ref. \cite{Belle} above and refs. \cite{ZhuKou}~\cite{IddirY4260} interpret $Y(4260)$ as a $c\bar{c}g$ hybrid. Amongst the possible interpretations of the recently discovered $XYZ$ mesons, hybrid option is commonly advocated; see refs. \cite{Godfrey, YuanForBelle}.

As mentioned above, it is because of the theoretical work on hybrids that we know how to interpret or not an exotic meson as a hybrid, though this type of work needs improvements both in terms of more specific predictions and clearer relationship with QCD. Available studies include, in addition to the ones mentioned in the above paragraph, those through the Flux-Tube Model~\cite{Ding Yan}, ~\cite{Isgur}-\cite{SwansonFluxTube}, QCD string model~\cite{Kalash08,Towers}, the Quark Model with a Constituent
Gluon~\cite{Swanson1998}-\cite{Yu.B.Yufryakov}, the Lattice QCD~\cite{Michael06}-\cite{Prelovsek2013}, QCD Sum Rules ~\cite{Berg12}-\cite{chen}, using Many-Body Coulomb Gauge Hamiltonian~\cite{Cotanch07}-\cite{Adam08}.

Testing models of QCD and our understanding of QCD effects, as well as interpreting the recent meson spectrum,
all point towards the significance of charmonium system. For example, Barnes says \cite{Barnestalk} ``charomonium system is an excellent laboratory for studyding non-perturbative effects such as confinement and gluonic excitations'' and ``identification of complete hybrid multiplet, especially $J^{PC}$ exotics, would be a crucial contribution to our
understanding of the dynamics of gluonic excitations.'' Ketzer says in ref. \cite{Ketzer} ``To avoid experimental difficulties in the light quark sector due to the high density of ordinary $q\bar{q'}$ states below 2.5 GeV$/c^2$, a search for hybrids in the less populated charmonium mass region is
expected to be very rewarding.'' This opinion is re-iterated by many others and most of them works on hybrids; see, for example,
 refs. \cite{Belle},\cite{ZhuKou}-\cite{YuanForBelle},~\cite{Kalash08,Iddir,12045425,Moir2013}~\cite{Luo Mei}-\cite{Dudek},~\cite{Prelovsek2013}-\cite{Harnett12},~\cite{Qiao,hernet13,Cotanch07,Cotanch}.

A possible approach to take advantage of all the theoretical approaches to understand hybrid mesons is to use the numbers generated by lattice simulations of QCD (for discretized configurations of quarks and antiquarks) to constrain form and parameters of a continuum model that is motivated directly or indirectly by QCD. Once that is done, we can compare the model with the results of actual hard experiments. The phenomenological implications we report result from fitting to the lattice result an explicitly written extension of the quark potential $V_{q\bar{q}}(\mathbf{r})$
whose use is $\emph{not}$ limited to finding properties of the ground state hybrid mesons; $(\mathbf{r})$
 is the relative vector a quark and antiquark.
 Like any potential model, the relation of the model to QCD can perhaps be most easily seen through what is termed the Born-Oppenheimer approach already used in hadronic physics in ref. \cite{morningstar} and later in
 refs. \cite{Nosheen11}~\cite{Braaten}:
  taking energy of the
 quickly adjusting gluonic field, found through the numerical lattice simulations of QCD, to be our adiabatic quark-antiquark potential
 $V_{q\bar{q}}(\mathbf{r})$, we first fit  its parameters.
  Then we solve the resulting
Schr$\ddot{o}$dinger equation for
a set of possible quantum numbers of quark anti-quark (for all cases) to calculate the meson
 energies and wave functions; for hybrid we used the additional term \cite{Nosheen11} in our potential and considered quantum numbers of the gluonic field as well. The result has been the directly testable implications for the hadronic physics we report.

In our earlier work~\cite{Nosheen11}, properties of conventional and hybrid charmonium mesons  were studied for radially ground state using an extended quark model. The investigation of radial excitations of mesons and hybrid mesons is of great interest  in hadron physics.
In refs. \cite{arxivbib,Liu PRD74} few radial excited states of conventional and hybrid charmonium mesons were
predicted, though not with any potential model; ref. \cite{Cotanch07} reports some orbital excitations of hybrid. In this
present work, we calculate the masses, sizes, and radial wave functions of ground and radial excited state conventional ($\eta, J/ \psi, h_c,$ $\chi_{o}$,$\chi_{1}$, $\chi_{2}$,....) and hybrid charmonium states such as
 $ 0^{--}, 0^{+ -}, 1^{- +}, 2^{+ -}, 3^{- +}...$ $J^{PC}$ states.
 This present work is the extension of our previous work~\cite{Nosheen11}. Using the radii
form factors~\cite{Ananthanarayan, Heglin}
energy shifts~\cite{S. I. Kruglov},~\cite{wang}, and magnetic polarizabilities~\cite{S. I.
Kruglov} can be found for conventional and hybrid
charmonium mesons. The
decay constants~\cite{Bhavin},~\cite{Qian decay} decay rates~\cite{Bhavin},~\cite{eichten}, and
differential cross sections~\cite{diff} for quarkonium states can be calculated using the radial wave function at the origin.

The paper is organized as follows. In the section II, the potential model used for conventional mesons is written. Then using this potential model, radial wave functions for the ground and radially excited state conventional charmonium mesons are found by numerically solving the Schr$\ddot{o}$dinger equation. The expressions used to find masses, root mean square radii, and radial wave functions at the origin for conventional charmonium mesons are also written
in this section. In section III, the potential model is written for hybrid mesons, and then we accordingly (i.e. for hybrids) redo  all the numerical work as done in section II. Results for the masses, root mean square radii
and square of radial wave functions at the origin for the radial and orbital ground and excited states of conventional and hybrid charmonium mesons are reported in section IV. Based on these results, we also include some results related to experimentally
measurable quantities. In section IV, XYZ- particles are related with different states of charmonium mesons based on similar mass and $J^{PC}$ states where available.
\qquad

\section*{II. Conventional charmonium mesons}

For the conventional mesons, we use the following quark anti-quark effective potential~\cite{charmonia05}
\begin{equation}
V_{q\overline{q}}(r)=\frac{-4\alpha _{s}}{3r}+br+\frac{32\pi \alpha _{s}}{
9m_{c}^{2}}(\frac{\sigma }{\sqrt{\pi }})^{3}e^{-\sigma ^{2}r^{2}}\overrightarrow{S}_{c}.\overrightarrow{S}_{
\overline{c}}+\frac{1}{m_{c}^{2}}[(\frac{2\alpha _{s}}{r^{3}}-\frac{b}{2r})
\overrightarrow{L}.\overrightarrow{S}+\frac{4\alpha _{s}}{r^{3}}T],
\label{P24}
\end{equation}
where the first term  is due to one gluon exchange with quark-gluon coupling
constant $\alpha _{s}$, second is linear confining potential
with string tension $b$, third term is the Gaussian-smeared hyperfine
interaction, and the last term is for the spin orbit
potential with
\begin{equation}
\overrightarrow{L}.\overrightarrow{S}=[J(J+1)-L(L+1)-S(S+1)]/2,
\end{equation}
and
\begin{equation}
<^3L_J\mid T \mid^3L_J>= \Bigg \{
\begin{array}{c}
-\frac{1}{6(2L+3)},J=L+1 \\
+\frac{1}{6},J=L \\
-\frac{L+1}{6(2L-1)},J=L-1.
\end{array}
\end{equation}
Here $L$ and $S$ are quantum numbers of the relative orbital angular momentum of  quark-antiquark and the total spin angular momentum of the meson respectively. The spin-orbit potential and the tensor term~\cite{charmonia05} are both zero for the angular $L=0$. The parameters used in this
potential for the charm quark and anti-quark are taken to be $\alpha
_{s}=0.5461$, $b=0.1425\text{ GeV}^{2}$, $\sigma =1.0946$ GeV, $m_{c}=1.4796$
GeV as in refs. \cite{Nosheen11,charmonia05}. These values are obtained from
the fit of the masses of 11 experimentally known $c\overline{c}$ states mentioned in the last column of Table 1 below. In the third term, $\overrightarrow{S}_{c}.\overrightarrow{S}_{\overline{c}}=\frac{S(S+1)}{2
}-\frac{3}{4}$,
$\mu $ is the reduced mass of the quark and antiquark, and $
m_{c}$ is the mass of the charm quark.
The values of quantum numbers $L$ and
$S$ which we choose for our study are reported in Table 1 below.
In the quark model, the characteristics of a conventional meson can be described by the wave function of the bound state of quark-antiquark and in the above  mentioned B.O. approximation this wave function  $U(r)=rR(r)$ can be found by solving the radial Schr$\ddot{\text{o}}$dinger
equation given as
\begin{equation}
U^{\prime \prime }(r)+2\mu (E-V(r)-\frac{L(L+1)}{2\mu r^{2}})U(r)=0.
\label{P23}
\end{equation}
$R(r)$ is the radial wave function and $r$ is the interquark distance. Here $E$ is the sum of kinetic and potential energy of quark-anti-quark system.
In  this non-relativistic approximation, especially justified for heavy mesons, the mass of a $c\overline{c}$ state is obtained after the addition of
constituent quark masses in the energy $E$. We  found the numerical solution of the Schr$\ddot{\text{o}}$dinger equation using the shooting method. Earlier~\cite{Nosheen11} we used this method to find the
numerical solutions for radially ground state ( i.e for $n=1$). Now we
extended this work of ref. \cite{Nosheen11} by finding the normalized
solutions for radially ground and excited states (i.e for $n=1,2,3,4,...$).
The numerical solutions were obtained by regularizing the $\frac{1}{r^{3}}$
term (of eq. (1)) by
adding a parameter $\epsilon $ to $r$ in the denominator whose value is taken small enough not to affect the equation for $r$ values away from the origin.  This remedy is obviously required    only for the mesons for which $L>0$.
  As a check on our calculation, we confirmed that our results for the
charmonium meson masses for ground and radially excited states agree with
the Table 1 of ref.~\cite{charmonia05}.
 \begin{figure}
\begin{center}
\epsfig{file=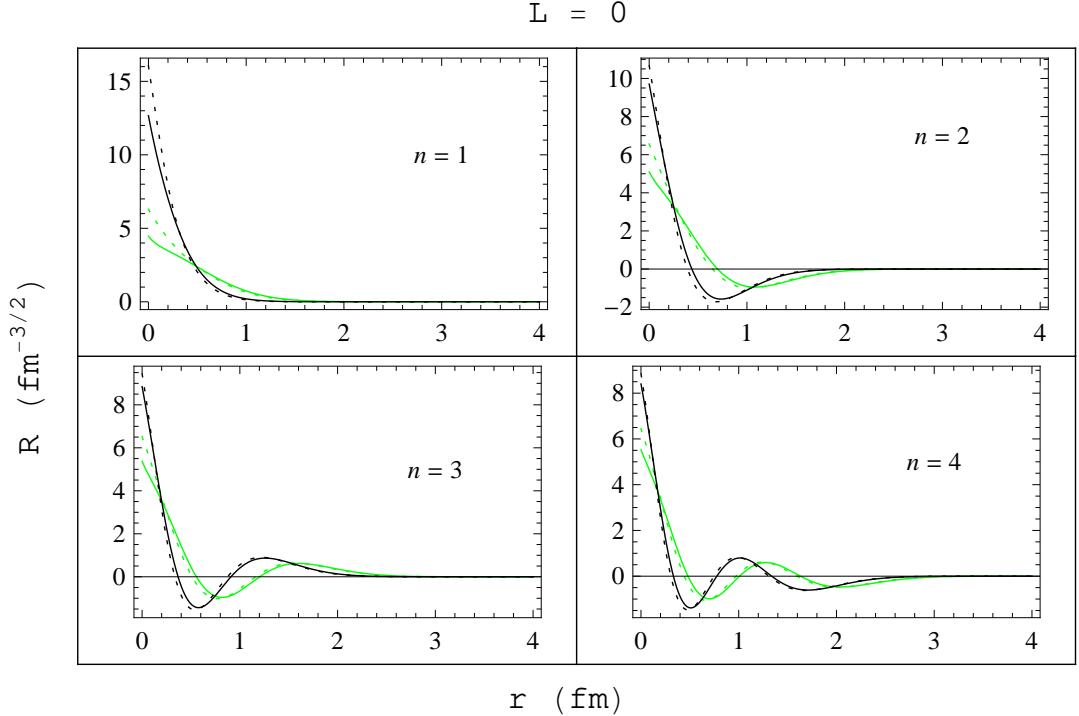,width=0.9\linewidth,clip=}\caption{Radial wave functions for radially ground and excited states of $\eta_c$ and $J/\psi$ mesons. Black color represents
mesons and green color represents hybrid mesons. Solid line curves are for $J/\psi$ and dotted curves are for $\eta_c$.}
\end{center}
\end{figure}
\begin{figure}
\begin{center}
\epsfig{file=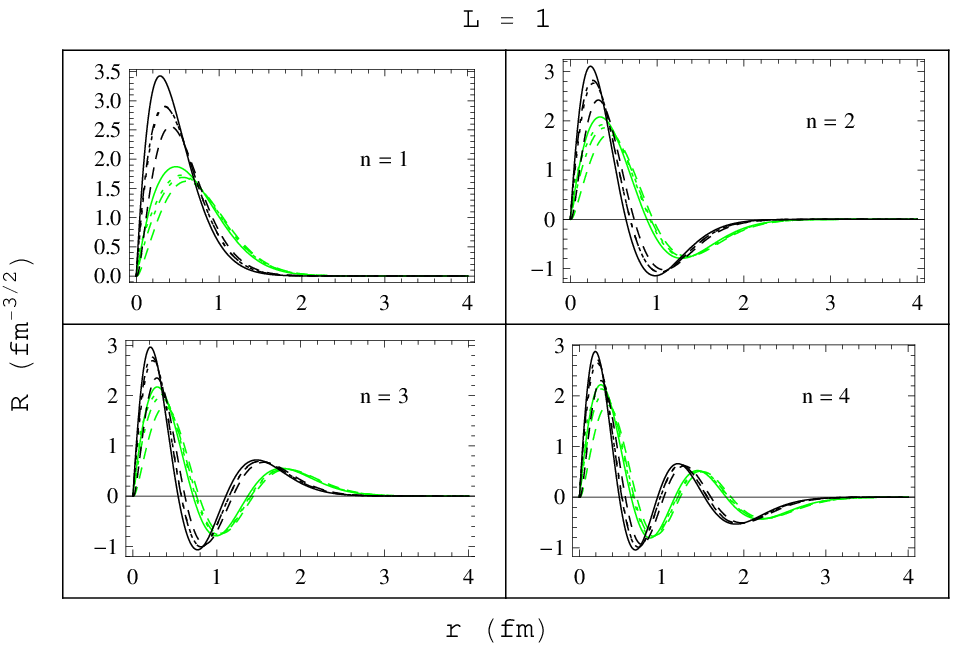,width=0.9\linewidth,clip=}\caption{Radial wave functions for radially ground and excited states of $\chi_{c_0}$, $\chi_{c_1}$, $\chi_{c_2}$, and $h_c$. Solid line curves are for $\chi_{c_0}$, dotted for $\chi_{c_1}$,
dashed for $\chi_{c_2}$, and dot dashed for $h_c$. Black color represents
mesons and green color represents hybrid mesons.}
\end{center}
\end{figure}
\begin{figure}
\begin{center}
\epsfig{file=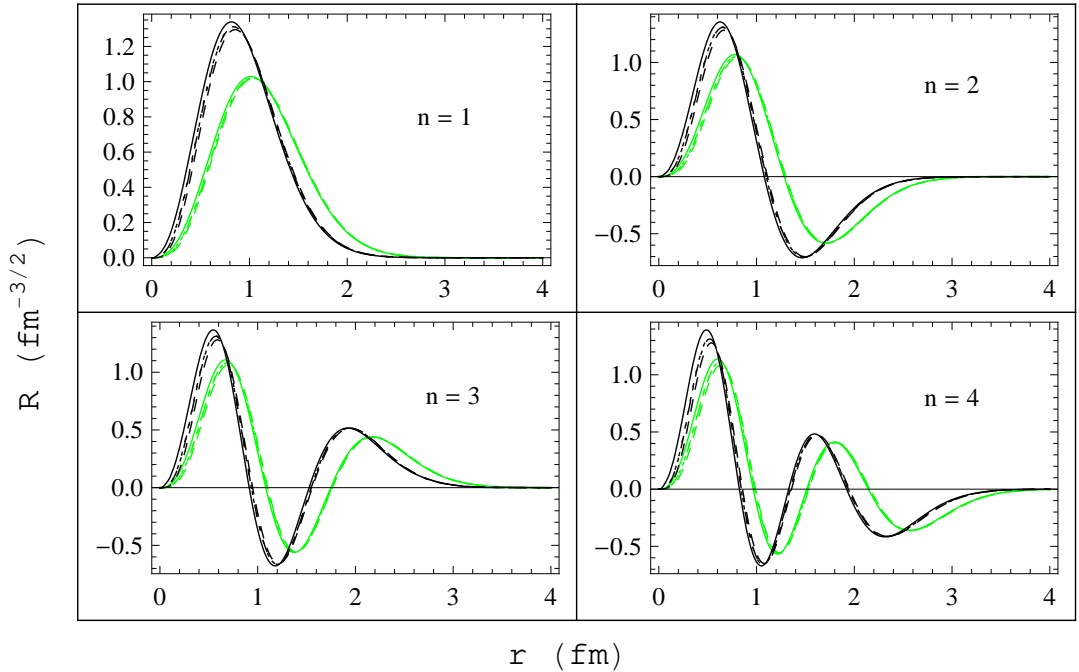,width=0.9\linewidth,clip=}\caption{Radial wave functions for radially ground and excited states of $\psi_1$, $\psi_{2}$, $\psi_{3}$ and $\eta_{c_2}$. Solid line curves are for $\psi_1$, dotted for $\psi_2$, dashed for $\psi_3$, and dot dashed for $\eta_{c_2}$. Black color represents mesons and green color represents hybrid mesons.}
\end{center}
\end{figure}
\begin{figure}
\begin{center}
\epsfig{file=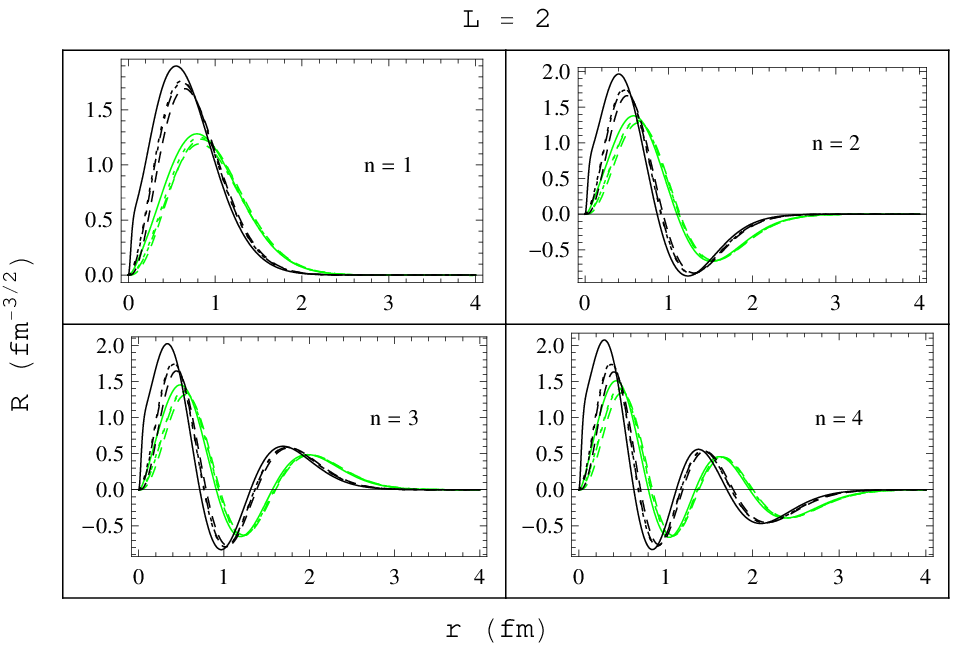,width=0.9\linewidth,clip=}\caption{Radial wave functions for radially ground and excited states of $\chi_{2}$, $\chi_{3}$, $\chi_{4}$ and $h_{c_3}$. Solid line curves are for $\chi_{2}$, dotted for $\chi_{3}$, dashed for $\chi_{4}$, and dot dashed for $h_{c_3}$. Black color represents mesons and green color represents hybrid mesons.}
\end{center}
\end{figure}
The black curves in Figs. 1 to 4 show the radial wave functions $R(r)$ of
charmonium mesons for different values of $n$, $L$, and $S$. Four
panels in each Fig. correspond to four values of $n=1$, $2$, $3$, and $4$
respectively and four black curves in each panel correspond to four
possible values of $J$ corresponding to each value of $L$, which are taken
0, 1, 2, and 3 in Figs. 1, 2, 3, and 4 respectively. The exception of two
black curves in each panel of Fig. 1 arise because for $L=0$ the quantum
number $J$ can have only two possible values. We noted that for $l>0$, $R(0)=0$ and $R(r)$  decrease exponentially at large inter
quark distances. The number of nodes in a radial wave function is equal to
$n-1$. It is also noted that peaks come closer to origin as we go
to higher radially excited states. And by increasing $L$, the wave
function's peak goes away from the origin.
But the wave functions are essentially insensitive to the $S$ values.  One possible reason is that for our heavy quarks the $1/m_{c}^{2}$ factor
(appearing in eq. \eqref{P24}) of the hyperfine term becomes very small. The normalized wave functions are then used to calculate root mean square radii and radial wave functions at origin using the following relations:
\begin{equation}
\sqrt{\langle r^{2}\rangle }=\sqrt{\int U^{\star }r^{2}Udr}.  \label{P25}
\end{equation}
\begin{equation}
R(0)=U^{\prime }(0)\text{ for }l=0.  \label{mid}
\end{equation}
Radial wave function at origin is used in many
applications of high energy physics as mentioned in section I.

\section*{III. Characteristics of Hybrid Charmonium mesons}

To describe hybrid mesons in BO approxiamtion, we used the static potential $
V_{q\overline{q}}^{h}(r)$ in place of $V_{q\overline{q}}(r)$ of eq. (\ref{P24}):
\begin{equation}
V_{q\overline{q}}^{h}(r)=V_{q\overline{q}}(r)+V_{g}(r)  \label{hypot}
\end{equation}

\noindent where $V_{g}(r)$ is the gluonic potential whose functional form varies with
the level of gluonic excitation. This potential and the corresponding
gluonic states are labeled by
  the magnitude ($\Lambda =0,1,2...$
 corresponding to greek letters $\Sigma, \Pi, \Delta, ...$) of the projection of total angular
momentum of gluons onto quark anti-quark axis and the behavior of
projection under the combined operation of charge conjugation and spatial
inversion. The states which are even (odd) under this operation are
represented by  a subscript $\eta =g(u)$ on the label. However, the $
\Sigma $ states also require the sign\ of parity under
reflection in the
plane perpendicular to quark anti-quark axis for their unique specification.
The sign of this parity is put in the superscript of the state's label. In
present work we study the hybrids in which the gluons are in the first
excited state $\Lambda =1$. This state is represented by the label $\Pi _{u}$. For
hybrid mesons the radial differential equation is given by
\begin{equation}
U^{\prime \prime }(r)+2\mu \left(E-V_{q\overline{q}}^{h}(r)-\frac{\left\langle
L_{q\overline{q}}^{2}\right\rangle }{2\mu r^{2}}\right)U(r)=0,  \label{hyrad}
\end{equation}

\noindent where squared quark anti-quark angular momentum $\left\langle L_{q\overline{q%
}}^{2}\right\rangle $ \cite{Kuti97,Juge99} is given by%
\begin{equation}
\left\langle L_{q\overline{q}}^{2}\right\rangle =L(L+1)-2\Lambda
^{2}+\left\langle J_{g}^{2}\right\rangle.
\end{equation}

\noindent For the $\Pi _{u}$ state, the squared gluon angular momentum $\left\langle
J_{g}^{2}\right\rangle =2$ and $\Lambda =1$~\cite{Kuti97} making
$-2\Lambda ^{2}+\langle J_{g}^{2}\rangle =0$.
The parity and charge parity of
hybrid meson are given by
\begin{equation}
P=\epsilon (-1)^{L+\Lambda +1},C=\epsilon \eta (-1)^{L+\Lambda +S},
\label{cp}
\end{equation}
\noindent where $\eta =-1$ and $\epsilon =\pm 1$ for $\Pi _{u}$ state\cite{Kuti97}. In the present
work we use following $V_{g}(r)$
\begin{equation}
V_{g}(r)=\frac{c}{r}+A\times exp^{-Br^{0.3723}}  \label{vg}
\end{equation}
\noindent where values of the constants $A=1.814$ and $B=0.066$ are fixed by
our earlier fit~\cite{Nosheen11} to the lattice data \cite{Kuti97} of the parameters of the effective potential form corresponding to $\Pi _{u}$ gluonic state. It is shown in ref. \cite{Nosheen11} that form of eq. (\ref{vg}) provides best fit to the lattice data.

Using the hybrid potential of eq. (\ref{hypot}) for $\Pi _{u}$ gluonic state in eq. (\ref{hyrad}), we calculated the masses and radial
wave functions of the hybrid mesons by using the same technique as employed
for conventional mesons (mentioned in section II). The resultant wave
functions are plotted in green color in Figs. 1-4 corresponding to the same
values of $n$, $L$, and $S$. These figures also show the comparison of the
conventional and hybrid meson radial wave functions. The shape of these radial wave functions is not much affected by the addition of the $V_g$ term for hybrids, though the values of masses are significantly increased  for the same values of $n$, $L$, and $S$.
\bigskip
\section*{IV. Results and Conclusions}

In previous work~\cite{Nosheen11} we calculated masses, root mean
square radii, and radial wave functions of $0^{+-},1^{-+}$, and $2^{+-}$ $%
J^{PC}$ states of the conventional and hybrid charmonium mesons. Now, considering the expected phenomenological challenges in forceable future, we
have extended this work~\cite{Nosheen11} by calculating these properties for
a rather comprehensive spectrum of conventional and hybrid charmonium mesons. We do this by including the radially excited $J^{PC}$ states not addressed in the previous study. In Table 1, our calculated masses are reported for the
ground and radially excited states of charmonium mesons along with the
experimental and theoretical predictions of the other's works. Our
calculated masses agree with the masses reported in
refs.~\cite{charmonia05, Ebert} as
well as the experimental values. In Table 2 the calculated masses of hybrid
charmonium mesons are reported for same values of $n$, $L$, and $S$ as used
for the conventional mesons. In order to distinguish the hybrids from
non-hybrids, we
suggest and use here a workable notation of using a superscript $h$ to the symbol of the conventional meson with the
same $n$, $L$, and $S$. These results show that for the same quantum numbers ($n$, $L$, and $S$) the mass of a
hybrid meson is significantly greater than the corresponding conventional meson. It is noted that $J^{PC}$ of each hybrid meson is also different from the
corresponding conventional meson for same $L$, and $S$. This difference arises
from the effect of the angular momentum of the gluonic field which contributes in the
former case. It is also noted that the gluonic potential $\Pi _{u}$
applied in this work allows two possible value of $\varepsilon $ in eq. (\ref{cp}).  As a result we obtain two degenerate hybrid states with opposite values
of $P$ and $C$. All the hybrids corresponding to $\varepsilon =1$ are
non-exotic, whereas exotic hybrid mesons are obtained for $\varepsilon =-1$
as shown in the Table 2. We find that the lightest hybrid charmonium state
has mass 4.0802 GeV and $J^{PC}=0^{++}(0^{--})$.
This result can be compared with the result reported in refs. \cite{Isgur1,Isgur4} that use
Flux tube model to predict that
the lowest state charmonium hybrid meson mass is approximately 4.2 GeV.
The similar result for the lowest state hybrid charmonium meson's mass (4.09
GeV) is predicted in ref. \cite{Iddir04} by using linear plus coulombic
potential model.

By comparing the experimental masses of various $X$, $Y$, $Z$ particles with
our calculated masses having same $J^{PC}$, we
suggest assigning the states calculated
by us to the experimentally observed particles as in Table 3. These results can help in experimentally recognizing hybrid mesons. In Table 4 we present the comparison of our results with other
theoretical studies of hybrid charmunium states in Flux tube model,
constitute glue model, and Lattic QCD. Root mean square radii and radial wave functions at origin for the ground
and radially excited states of conventional and hybrid charmonium mesons are
reported in Table 5 and 6 respectively. These results show that with the
same quantum numbers ($n$, $L$, and $S$) root mean square radii of hybrid mesons
are greater than conventional mesons. It is also noted that radii of hybrid
mesons like conventional mesons increase with radial and angular
excitations. Table 4 and 5 show that
$|R(0)|^{2}$ is non-zero only for $S$ states and decrease with the radial
excitation. As scalar form factors~\cite{Ananthanarayan}, energy shifts, and polarizabilites ~\cite{S. I. Kruglov}  depend on the root mean square radii, we
predict a significant change in the values of these quantities for a hybrid meson as compared to the corresponding conventional meson for the same quantum numbers. Thus it is highly interesting to compare our results with experimental findings of conventional and exotic mesons.

\begin{table}\caption{Masses of ground and radially excited state charmonium mesons. Our calculated masses are rounded to 0.0001 GeV.}
\begin{center}
\begin{tabular}{|c|c|c|c|c|c|c|c|c|c|}
\hline
n & Meson & $L$ & $S$ & J& $J^{PC}$& Our calculated & Theor. mass~\cite{charmonia05} & Exp. mass \\
& & & & & & mass & with NR potential & \\
& & & &  & & & model & \\ \hline
& & & &  & &  \textrm{GeV} & \textrm{GeV} & \textrm{GeV} \\ \hline
1S & $ \eta_{c} (1 ^1S_0)$ & 0 & 0 & 0 &$0^{-+}$ &  2.9816 & 2.982 &$2.9810\pm 0.0011$~\cite{pdg} \\
 &$J/\psi (1 ^3S_1)$& $0$ & $1$ &1 &$1^{--}$ &  $3.08999$ & $3.090$ & $3.096916\pm 0.000011$~\cite{pdg} \\ \hline
2 S & $ \eta'_{c} (2 ^1S_0)$ & 0 & 0 & 0&$0^{-+}$ &  3.6303 & 3.630 & $3.6389\pm 0.0013 $~\cite{pdg} \\
 & $J/\psi (2 ^3S_1)$& $0$ & $1$ & 1 &$ 1^{--}$ & $3.6718$ & 3.672 & $3.6861 _{-0.000014}^{+0.000012} $ ~\cite{pdg}\\ \hline
3S & $ \eta_{c} (3 ^1S_0)$ & 0 & 0  & 0&$0^{-+}$ & 4.0432 & 4.043 &--- \\
 & $J/\psi (3 ^3S_1)$& $0$ & $1$ &1 &$1^{--}$ &  $4.0716$ & 4.072 & $4.040 \pm 10$~\cite{charmonia05}\\ \hline
4 S & $ \eta_{c} (4 ^1S_0)$ & 0 & 0 &0 &0$^{-+}$ &  4.3837 & 4.384 &--- \\
 & $J/\psi (4 ^3S_1)$& $0$ & $1$ &1 &$1^{--}$ &  $4.4061$ & 4.406 & $4.415 \pm 6$~\cite{charmonia05} \\ \hline
5 S& $ \eta_{c} (5 ^1S_0)$ & 0 & 0 &0 &0$^{-+}$ &  4.6850 & --- &--- \\
 & $J/\psi (5 ^3S_1)$& $0$ & $1$ &1 &$1^{--}$ & $4.7038$ & --- &  \\ \hline
6 S& $ \eta_{c} (6 ^1S_0)$ & 0 & 0 &0 &0$^{-+}$ &  4.9604 &--- &--- \\
 & $J/\psi (6 ^3S_1)$& $0$ & $1$ &1 &$1^{--}$ &  $4.9769$ & --- & ---\\ \hline
1P & $ h_{c} (1 ^1P_1) $&  1 & $0$ & 1 &$1^{+-}$ &  $3.5156$ & $3.516$ & $3.52541\pm 0.00016$~\cite{pdg}\\
 & $\chi_{0} (1 ^3P_0)$ & $1$ & $1$ & 0 & $0^{++}$ & $3.4233$& $3.424$ & $3.41475\pm0.00031$ ~\cite{pdg}\\
 & $\chi_{1} (1 ^3P_1)$ & $1$ & $1$ &1 &$1^{++}$ &  $3.5005$ & $3.505$ & $3.51066 \pm 0.00007$~\cite{pdg} \\
 & $\chi_{2} (1 ^3P_2)$ & $1$ & $1$ &2 &$2^{++}$ &  $3.5490$ & $3.556$ & $3.55620 \pm 0.00009$~\cite{pdg} \\ \hline
2 P& $h_{c} (2 ^1P_1) $&  1 & $0$ &1 & $1^{+-}$&  $3.9336$ & 3.934 &--- \\
 & $\chi_{0} (2 ^3P_0)$ & $1$ &1&0 &$ 0^{++}$ & $3.8715$& 3.852 & --- \\
 & $\chi_{1} (2 ^3P_1)$ & $1$ & $1$ &1 &$1^{++}$ &  $3.9203$ & 3.925 & ---\\
 & $\chi_{2} (2 ^3P_2)$ & $1$ & $1$ & 2&$2^{++}$ &  $3.9648 $& 3.972 &$3.9272 \pm 0.0026$~\cite{pdg} \\ \hline
3 P & $h_{c} (3 ^1P_1) $&  1 & $0$ & 1 &$1^{+-}$ &  $4.2793$ & 4.279 &--- \\
 & $\chi_{0} (3 ^3P_0)$ & $1$ &1 &0 & $0^{++}$ & $4.2295$& 4.202 & ---\\
 & $\chi_{1} (3 ^3P_1)$ & $1$ & $1$ &1 &$1^{++}$ &  $4.2663$& 4.271 & ---\\
 & $\chi_{2} (3 ^3P_2)$ & $1$ & $1$ &2 &$2^{++}$ &  $4.3093$& 4.317 & ---\\ \hline
4 P& $h_{c} (4 ^1P_1) $&  1 & $0$ & 1 &$1^{+-}$ &  $4.5851$ & --- &--- \\
 & $\chi_{0} (4 ^3P_0)$ & $1$ & $1$ &0 &$0^{++}$ &  $4.5424$& --- & ---\\
 & $\chi_{1} (4 ^3P_1)$ & $1$ & $1$ &1  &$1^{++}$ & $4.5720$& --- & ---\\
 & $\chi_{2} (4 ^3P_2)$ & $1$ & $1$ & 2&$ 2^{++}$&  $4.6141$& --- & ---\\ \hline
5 P & $h_{c} (5 ^1P_1) $&  1 & $0$ & 1 &$1^{+-}$ &  $4.8644$ & --- &---\\
 & $\chi_{0} (5 ^3P_0)$ & $1$ & $1$ &0 &$0^{++}$ &  $4.8264$& --- & ---\\
 & $\chi_{1} (5 ^3P_1)$ & $1$ & $1$ &1  &$1^{++}$ & $4.8512$& --- & ---\\
 & $\chi_{2} (5 ^3P_2)$ & $1$ & $1$ & 2&$ 2^{++}$&  $4.8926$& --- & ---\\ \hline
6 P & $h_{c} (6 ^1P_1) $&  1 & $0$ & 1 &$1^{+-}$ &  $5.1244$ & --- &--- \\
 & $\chi_{0} (6 ^3P_0)$ & $1$ & $1$ &0 &$0^{++}$ &  $5.0898$& --- & ---\\
 & $\chi_{1} (6 ^3P_1)$ & $1$ & $1$ &1  &$1^{++}$ & $5.111$& --- & ---\\
 & $\chi_{2} (6 ^3P_2)$ & $1$ & $1$ & 2&$ 2^{++}$& $5.15198$& --- & ---\\ \hline
1 D & $ \eta_{c2} (1 ^1D_2)$ & 2 & 0 & 2 &$2^{-+}$ & 3.7994 & 3.799 &--- \\
 &$\psi (1 ^3D_1)$& 2 & 1 &1 &$1^{--}$ &  3.7805 & $3.785$ &  $3.7699\pm 0.0025$ ~\cite{charmonia05}  \\
 &$\psi_{2} (1 ^3D_2)$& 2 & 1 & 2 &$2^{--}$ &  3.8002 & $3.800$ & ---\\
 &$\psi_{3} (1 ^3D_3)$& 2 & 1 &3 &$3^{--}$ & 3.8053 & $3.806$ &---\\ \hline
\end{tabular}
\end{center}
\end{table}
\begin{table}
\begin{center}
\begin{tabular}{|c|c|c|c|c|c|c|c|c|c|}
\hline
n & Meson & $L$ & $S$ & J& $J^{PC}$& Our calculated & Theor. mass~\cite{charmonia05} & Exp. mass \\
& & & & & & mass & with NR potential & \\
& & & &  & & & model & \\ \hline
& & & &  & &  \textrm{GeV} & \textrm{GeV} & \textrm{GeV} \\ \hline
2 D & $ \eta_{c2}(2 ^1D_2)$ & 2 & 0 & 2 &$2^{-+}$ & 4.1576 & 4.158 &--- \\
 &$\psi (2 ^3D_1)$& 2 & 1 &1 &$1^{--}$ &  4.1363 & $4.142$ & $4.159\pm 0.020$~\cite{charmonia05} \\
 &$\psi_{2} (2 ^3D_2)$& 2 & 1 & 2 &$2^{--}$ &  4.1580 & $4.158$ &--- \\
 &$\psi_{3} (2 ^3D_3)$& 2 & 1 &3 &$3^{--}$ &  4.1655 & $4.167$ & ---\\ \hline
3 D & $ \eta_{c2}(3 ^1D_2)$ & 2 & 0 & 2 &$2^{-+}$ & 4.4718 &--- & ---\\
 &$\psi (3 ^3D_1)$& 2 & 1 &1 &$1^{--}$ &  4.4492 &--- &--- \\
 &$\psi_{2} (3 ^3D_2)$& 2 & 1 & 2 &$2^{--}$ &  4.4719 & ---&--- \\
 &$\psi_{3} (3 ^3D_3)$& 2 & 1 &3 &$3^{--}$ &  4.4810 &--- &--- \\ \hline
4 D & $ \eta_{c2}(4 ^1D_2)$ & 2 & 0 & 2 &$2^{-+}$ & 4.7574 & ---&--- \\
 &$\psi (4 ^3D_1)$& 2 & 1 &1 &$1^{--}$ &  4.7339 &--- &--- \\
 &$\psi_{2} (4 ^3D_2)$& 2 & 1 & 2 &$2^{--}$ &  4.7573 &--- &--- \\
 &$\psi_{3} (4 ^3D_3)$& 2 & 1 &3 &$3^{--}$ &  4.7675 &--- &---\\ \hline
5 D & $ \eta_{c2}(5 ^1D_2)$ & 2 & 0 & 2 &$2^{-+}$ & 5.0223 &--- &--- \\
 &$\psi (5 ^3D_1)$& 2 & 1 &1 &$1^{--}$ &  4.9984 &--- &---  \\
 &$\psi_{2} (5 ^3D_2)$& 2 & 1 & 2 &$2^{--}$ &  5.0221 &---&--- \\
 &$\psi_{3} (5 ^3D_3)$& 2 & 1 &3 &$3^{--}$ &  5.0331 &--- &--- \\ \hline
1 F& $h_{c3}(1 ^1F_3) $&  3 & $0$ & 3 &$3^{+-}$ &  $4.0256$ & 4.026 &--- \\
 & $\chi_2 (1 ^3F_2)$ & $3$ & $1$ &2 &$2^{++}$ &  $4.0283$& 4.029 & ---\\
 & $\chi_3(1 ^3F_3)$ & $3$ & $1$ &3  &$3^{++}$ & $4.0287$& 4.029 & ---\\
 & $\chi_4(1 ^3F_4)$ & $3$ & $1$ & 4 &$ 4^{++}$&  $4.0212$& 4.021 & ---\\ \hline
2 F& $h_{c3}(2 ^1F_3) $&  3 & $0$ & 3 &$3^{+-}$ &  $4.3499$ & 4.350 &--- \\
 & $\chi_2(2 ^3F_2)$ & $3$ & $1$ &2 &$2^{++}$ &  $4.3494$& 4.351& ---\\
 & $\chi_3(2 ^3F_3)$ & $3$ & $1$ &3  &$3^{++}$ & $4.3522$& 4.352 & ---\\
 & $\chi_4(2 ^3F_4)$ & $3$ & $1$ & 4 &$ 4^{++}$&  $4.3476$& 4.348& ---\\ \hline
3 F& $h_{c3}(3 ^1F_3) $&  3 & $0$ & 3 &$3^{+-}$ &  $4.6429$ & --- &--- \\
 & $\chi_2(3 ^3F_2)$ & $3$ & $1$ &2 &$2^{++}$ &  $4.6403$& --- & ---\\
 & $\chi_3(3 ^3F_3)$ & $3$ & $1$ &3  &$3^{++}$ & $4.6448$& --- & ---\\
 & $\chi_4(3 ^3F_4)$ & $3$ & $1$ & 4 &$ 4^{++}$&  $4.6422$& --- & ---\\ \hline
4 F& $h_{c3}(4 ^1F_3) $&  3 & $0$ & 3 &$3^{+-}$ &  $4.9137$ & --- &--- \\
 & $\chi_2(4^3F_2)$ & $3$ & $1$ &2 &$2^{++}$ &  $4.9095$& --- & ---\\
 & $\chi_3(4 ^3F_3)$ & $3$ & $1$ &3  &$3^{++}$ & $4.9153$& --- & ---\\
 & $\chi_4(4 ^3F_4)$ & $3$ & $1$ & 4 &$ 4^{++}$&  $4.9141$& --- & ---\\ \hline
1 G& $ \eta_{c4}(1 ^1G_4)$ & 4 & 0 & 4 &$4^{-+}$ & 4.2246 & 4225 & ---\\
 &$\psi_3 (1 ^3G_3)$& 4 & 1 &3 &$3^{--}$ &  4.23635 & 4.237 &--- \\
 &$\psi_4 (1 ^3G_4)$& 4 & 1 & 4 &$4^{--}$ & 4.22821 & 4.228 &--- \\
 &$\psi_5 (1 ^3G_5)$& 4 & 1 &5 &$5^{--}$ & 4.2142 & 4.214 &--- \\ \hline
\end{tabular}
\end{center}
\end{table}

\begin{table}\caption{Our calculated
masses of $c\overline{c}$ hybrid charmonium mesons.}
\begin{center}
\begin{tabular}{|c|c|c|c|c|c|c|c|}
\hline
n & Meson & $L$ & $S$ & J& \multicolumn{2}{|c|}{$J^{PC}$}& Our calculated  \\
& & & & & $\varepsilon=1$& $\varepsilon=-1$ & mass \\
 & & & &  & & &  \textrm{GeV} \\ \hline
1S & $ \eta^{h}_{c} (1 ^1S_0)$ & 0 & 0 & 0 &$0^{++}$ & $0^{--}$ & 4.0802  \\
 &$ J/\psi^{h} (1 ^3S_1)$& $0$ & $1$ &1 &$1^{+-}$ & $1^{-+}$ & $4.1063$  \\ \hline
2 S & $ \eta^{h}_{c} (2 ^1S_0)$ & 0 & 0 & 0&$0^{++}$ & $0^{--}$ & 4.3820 \\
 & $ J/\psi^{h} (2 ^3S_1)$& $0$ & $1$ & 1 &$ 1^{+-}$ & $1^{-+}$ & $4.4084$ \\ \hline
3S & $ \eta^{h}_{c} (3 ^1S_0)$ & 0 & 0  & 0&$0^{++}$ & $0^{--}$ & 4.6616  \\
 & $ J/\psi^{h} (3 ^3S_1)$& $0$ & $1$ &1 &$1^{+-}$ & $1^{-+}$ &  $4.6855$ \\ \hline
4 S & $ \eta^{h}_{c} (4 ^1S_0)$ & 0 & 0 &0 &$0^{++}$ & $0^{--}$ &  4.9223  \\
 & $ J/\psi^{h} (4 ^3S_1)$& $0$ & $1$ &1 &$1^{+-}$ & $1^{-+}$ & $4.9438$  \\ \hline
5 S& $ \eta^{h}_{c} (5 ^1S_0)$ & 0 & 0 & 0 &$0^{++}$ & $0^{--}$ & 5.1683  \\
 & $ J/\psi^{h} (5 ^3S_1)$& $0$ & $1$ &1 &$1^{+-}$ & $1^{-+}$ & $5.1876$  \\ \hline
6 S& $ \eta^{h}_{c} (6 ^1S_0)$ & 0 & 0 &0 &$0^{++}$ & $0^{--}$ & 5.4021  \\
 & $ J/\psi^{h} (6 ^3S_1)$& $0$ & $1$ &1 &$1^{+-}$ & $1^{-+}$ & $5.4197$  \\ \hline
1P & $ h^{h}_{c} (1 ^1P_1) $&  1 & $0$ & 1 &$1^{--}$ & $1^{++}$ & $4.2678$ \\
 & $ \chi^{h}_{0} (1 ^3P_0)$ & $1$ & $1$ & 0 & $0^{-+}$ & $0^{+-}$ & $4.22996$ \\
 & $ \chi^{h}_{1} (1 ^3P_1)$ & $1$ & $1$ &1 &$1^{-+}$ & $1^{+-}$ & $4.2664$ \\
 & $ \chi^{h}_{2} (1 ^3P_2)$ & $1$ & $1$ &2 &$2^{-+}$ & $2^{+-}$ & $4.2738$ \\ \hline
2 P& $ h^{h}_{c} (2 ^1P_1) $&  1 & $0$ &1 & $1^{--}$& $1^{++}$ & $4.5552$ \\
 & $ \chi^{h}_{0} (2 ^3P_0)$ & $1$ &1&0 &$ 0^{-+}$ & $0^{+-}$ &$4.5030$ \\
 & $ \chi^{h}_{1} (2 ^3P_1)$ & $1$ & $1$ &1 &$1^{-+}$ & $1^{+-}$ & $4.5511$ \\
 & $ \chi^{h}_{2} (2 ^3P_2)$ & $1$ & $1$ & 2&$2^{-+}$ & $2^{+-}$ & $4.5653$ \\ \hline
3 P & $ h^{h}_{c} (3 ^1P_1) $&  1 & $0$ & 1 &$1^{--}$ & $1^{++}$ & $4.8210$  \\
 & $ \chi^{h}_{0} (3 ^3P_0)$ & $1$ &1 &0 & $0^{-+}$ & $0^{+-}$ & $4.7627$ \\
 & $ \chi^{h}_{1} (3 ^3P_1)$ & $1$ & $1$ &1 &$1^{-+}$ & $1^{+-}$ &  $4.8149$\\
 & $ \chi^{h}_{2} (3 ^3P_2)$ & $1$ & $1$ &2 &$2^{-+}$ & $2^{+-}$ & $4.8337$ \\ \hline
4 P& $ h^{h}_{c} (4 ^1P_1) $&  1 & $0$ & 1 &$1^{--}$ & $1^{++}$ & $5.0707$  \\
 & $ \chi^{h}_{0} (4 ^3P_0)$ & $1$ & $1$ &0 &$0^{-+}$ & $0^{+-}$ & $5.0101$ \\
 & $ \chi^{h}_{1} (4 ^3P_1)$ & $1$ & $1$ &1  &$1^{-+}$ & $1^{+-}$ & $5.0629$\\
 & $ \chi^{h}_{2} (4 ^3P_2)$ & $1$ & $1$ & 2&$ 2^{-+}$& $2^{+-}$ & $5.0852$ \\ \hline
5 P & $ h^{h}_{c} (5 ^1P_1) $&  1 & $0$ & 1 &$1^{1--}$ & $1^{++}$ & $5.3076$ \\
 & $ \chi^{h}_{0} (5 ^3P_0)$ & $1$ & $1$ &0 &$0^{-+}$ & $0^{+-}$ &  $5.2466$ \\
 & $ \chi^{h}_{1} (5 ^3P_1)$ & $1$ & $1$ &1  &$1^{-+}$ & $1^{+-}$ & $5,2983$ \\
 & $ \chi^{h}_{2} (5 ^3P_2)$ & $1$ & $1$ & 2&$ 2^{-+}$& $2^{+-}$ & $5.3233$ \\ \hline
6 P & $ h^{h}_{c} (6 ^1P_1) $&  1 & $0$ & 1 &$1^{--}$ & $1^{++}$ & $5.5340$ \\
 & $ \chi^{h}_{0} (6 ^3P_0)$ & $1$ & $1$ &0 &$0^{-+}$ & $0^{+-}$ & $5.4734$ \\
 & $ \chi^{h}_{1} (6 ^3P_1)$ & $1$ & $1$ &1  &$1^{-+}$ & $1^{+-}$ & $5.5234$ \\
 & $ \chi^{h}_{2} (6 ^3P_2)$ & $1$ & $1$ & 2&$ 2^{-+}$& $2^{+-}$ & $5.55070$ \\ \hline
1 D & $ \eta_{c2} (1 ^1D_2)$ & 2 & 0 & 2 &$2^{++}$ & $2^{--}$ & 4.4222 \\
 &$\psi^{h} (1 ^3D_1)$& 2 & 1 &1 &$1^{+-}$ & $1^{-+}$ & 4.4220 \\
 &$\psi^{h}_{2} (1 ^3D_2)$& 2 & 1 & 2 &$2^{+-}$ & $2^{-+}$ &4.4251 \\
 &$\psi^{h}_{3} (1 ^3D_3)$& 2 & 1 &3 &$3^{+-}$ & $3^{-+}$ & 4.4197 \\ \hline
2 D & $ \eta^{h}_{c2} (2 ^1D_2)$ & 2 & 0 & 2 &$2^{++}$ & $2^{--}$ & 4.6982 \\
 &$\psi^{h} (2 ^3D_1)$& 2 & 1 &1 &$1^{+-}$ & $1^{-+}$ &  4.6932  \\
 &$\psi^{h}_{2} (2 ^3D_2)$& 2 & 1 & 2 &$2^{+-}$ & $2^{-+}$ & 4.7002 \\
 &$\psi^{h}_{3} (2 ^3D_3)$& 2 & 1 &3 &$3^{+-}$ & $3^{-+}$ &  4.6981 \\ \hline
\end{tabular}
\end{center}
\end{table}
\begin{table}
\begin{center}
\begin{tabular}{|c|c|c|c|c|c|c|c|c|}
\hline
n & Meson & $L$ & $S$ & J& \multicolumn{2}{|c|}{$J^{PC}$}& Our calculated   \\
& & & & & $\varepsilon=1$& $\varepsilon=-1$ & mass \\
& & & & & & &  \textrm{GeV} \\ \hline
3 D & $ \eta^{h}_{c2} (3 ^1D_2)$ & 2 & 0 & 2 &$2^{++}$ & $2^{--}$ & 4.9554 \\
 &$\psi^{h} (3 ^3D_1)$& 2 & 1 &1 &$1^{+-}$ & $1^{-+}$ & 4.9469 \\
 &$\psi^{h}_{2} (3 ^3D_2)$& 2 & 1 & 2 &$2^{+-}$ & $2^{-+}$ & 4.9568\\
 &$\psi^{h}_{3} (3 ^3D_3)$& 2 & 1 &3 &$3^{+-}$ & $3^{-+}$ & 4.9572 \\ \hline
4 D & $ \eta^{h}_{c2} (4 ^1D_2)$ & 2 & 0 & 2 &$2^{++}$ & $2^{--}$ & 5.19828 \\
 &$\psi^{h} (4 ^3D_1)$& 2 & 1 &1 &$1^{+-}$ & $1^{-+}$ & 5.18685 \\
 &$\psi^{h}_{2} (4 ^3D_2)$& 2 & 1 & 2 &$2^{+-}$ & $2^{-+}$ & 5.19928  \\
 &$\psi^{h}_{3} (4 ^3D_3)$& 2 & 1 &3 &$3^{+-}$ & $3^{-+}$ & 5.20141  \\ \hline
5 D & $ \eta^{h}_{c2} (1 ^1D_2)$ & 2 & 0 & 2 &$2^{++}$ & $2^{--}$ & 5.42954 \\
 &$\psi^{h} (5 ^3D_1)$& 2 & 1 &1 &$1^{+-}$ & $1^{-+}$ & 5.41568 \\
 &$\psi^{h}_{2} (5 ^3D_2)$& 2 & 1 & 2 &$2^{+-}$ & $2^{-+}$ & 5.43025 \\
 &$\psi^{h}_{3} (5 ^3D_3)$& 2 & 1 &3 &$3^{+-}$ & $3^{-+}$ & 5.43379  \\ \hline
1 F & $h^{h}_{c3}(1 ^1 F_3) $&  3 & $0$ & 3 &$3^{--}$ & $3^{++}$ &$4.57002$ \\
 & $\chi^{h}_{2}(1 ^3 F_2)$ & $3$ & $1$ &2 &$2^{-+}$ & $2^{+-}$ & 4.57886 \\
 & $\chi^{h}_{3}(1 ^3 F_3)$ & $3$ & $1$ &3 &$3^{-+}$ & $3^{+-}$ & $4.5735$ \\
 & $\chi^{h}_{4}(1 ^3 F_4)$ & $3$ & $1$ & 4&$ 4^{-+}$& $ 4^{+-}$& $4.5623$ \\ \hline
2 F & $h^{h}_{c3}(2 ^1 F_3) $&  3 & $0$ & 3 &$3^{--}$ & $3^{++}$ & $4.835$  \\
 & $\chi^{h}_{2}(2 ^3 F_2)$ & $3$ & $1$ &2 &$2^{-+}$ & $2^{+-}$ & 4.8409 \\
 & $\chi^{h}_{3}(2 ^3 F_3)$ & $3$ & $1$ &3 &$3^{-+}$ & $3^{+-}$ & $4.8379$ \\
 & $\chi^{h}_{4}(2 ^3 F_4)$ & $3$ & $1$ & 4&$ 4^{-+}$& $ 4^{+-}$& $4.8292$ \\ \hline
3 F & $h^{h}_{c3}(3 ^1 F_3) $&  3 & $0$ & 3 &$3^{--}$ & $3^{++}$ & $5.084$ \\
 & $\chi^{h}_{2}(3 ^3 F_2)$ & $3$ & $1$ &2 &$2^{-+}$ & $2^{+-}$ & 5.08763  \\
 & $\chi^{h}_{3}(3 ^3 F_3)$ & $3$ & $1$ &3 &$3^{-+}$ & $3^{+-}$ &  $5.0864$ \\
 & $\chi^{h}_{4}(3 ^3 F_4)$ & $3$ & $1$ & 4&$ 4^{-+}$& $ 4^{+-}$& $5.0798$ \\ \hline
4 F & $h^{h}_{c3}(4 ^1 F_3) $&  3 & $0$ & 3 &$3^{--}$ & $3^{++}$ & $5.3203$  \\
 & $\chi^{h}_{2}(4 ^3 F_2)$ & $3$ & $1$ &2 &$2^{-+}$ & $2^{+-}$ & 5.3222 \\
 & $\chi^{h}_{3}(4 ^3 F_3)$ & $3$ & $1$ &3 &$3^{-+}$ & $3^{+-}$ &  $5.3224$ \\
 & $\chi^{h}_{4}(4 ^3 F_4)$ & $3$ & $1$ & 4&$ 4^{-+}$& $ 4^{+-}$& $5.3173$ \\ \hline
\end{tabular}
\end{center}
\end{table}

\begin{table}\caption{Our assignments based on the mass equivalence of $J^{PC}$ and mass (i.e. comparison of our predicted
masses with the experimental masses.)}
\begin{center}
\begin{tabular}{|c|c|c|c|}
\hline
 Meson & $J^{PC}$ & Experimental mass & Assignments \\ \hline
 & & GeV &  \\ \hline
$\chi_{c_2}(2 P)$ & $2^{++}$ &$3.9272 \pm 2.6$ ~\cite{pdg} & $\chi_2(1 ^3 P_2)$ \\ \hline
$h_c(1 ^1P_1)$ & $1^{+-}$ & $3.52538 \pm 0.11$ ~\cite{pdg} & $h_c(1 ^1 P_1)$ \\ \hline
$X(3872)$ & $1^{++}$ & $3.87168 \pm 0.17$ ~\cite{pdg} & $ \chi_1 (2 ^3 P_1)$ \\
 & $?^{? +}$ &  & $\chi_0(2 ^3 P_0)$ , $\chi_1(2 ^3 P_1)$ , $\chi_2 (2 ^3 P_2)$, $\eta_{c2}(1 ^1 D_2)$ \\ \hline
 $X(4260)$ & $1^{--}$ & $4.250\pm 9$~\cite{pdg}  & $ \psi(2 ^3 D_1)$, $ \psi(3 ^3 D_1)$, $ h^h_c(1 ^1 P_1)$ \\ \hline
 $X(3915)$ & $0^{++}$ & $3.918.4 \pm 1.9$~\cite{pdg}  & $ \chi_0(2 ^3 P_0)$ \\
  & $?^{?+}$ & & $ \eta_c(3 ^1 S_0)$, $\chi_0(2 ^3 P_0)$, $\chi_1( 2 ^3 P_1)$, $\chi_2( 2 ^3 P_2)$ \\ \hline
  $X(4360)$ & $1^{--}$ & $4.361\pm 13$~\cite{pdg} & $\psi( 2 ^3 D_1)$, $ \psi(3 ^3 D_1)$, $J/\psi( 4 ^3 S_1)$ \\
   $Y(4360)$ & & & \\ \hline
   $X(4660)$ & $1^{--}$ &$4.664 \pm 12$~\cite{pdg}  & $ J/\psi(5 ^3 S_1)$, $ J/\psi(4 ^3 S_1)$, $ \psi^h(4 ^3 D_1)$ \\
  &  & & $ h^h_c(2 ^1 P_1)$, $ h^h_c(3 ^1 P_1)$\\ \hline
   $X(3940)$ & $1^{--}$ & $3.942^{+9}_{-8}$~\cite{1307} & $ J/\psi(3 ^3 S_1)$ \\ \hline
    $X(4350)$ & $1^{--}$ & $4.3506 \pm_{-5.1} ^{+4.6}$~\cite{1307},~\cite{Belle} & $ J/\psi(4 ^3 S_1)$, $ \psi(3 ^3 D_1)$, $ h^h_c(1 ^1 P_1)$ \\ \hline
   $X(4008)$ & $1^{--}$ & $4.008^{+121}_{- 49}$~\cite{1307} & $\psi( 2 ^3 D_1)$ \\ \hline
   $X(3940)$ & $1^{++}$ & ---- & $ \chi_0(2 ^3 P_0)$ \\ \hline
   $X(4630)$ & $1^{--}$ & $4.634_{-11}^{+9}$~\cite{1307}& $ J/\psi(5 ^3 S_1)$, $ \psi(4 ^3 D_1)$, $h^h_c( 2 ^1 P_1)$ \\
   $Y(4630)$ &  &  & \\ \hline
   $Y(4350)$ & $0^{++}$ & $4.3506^{+4.6}_{-5.1}$~\cite{1307} & $ \chi_0(3 ^3 P_0)$, $ \eta^h_c(2 ^1 S_0)$ \\
    & $2 ^{++}$ & $4.3506^{+4.6}_{-5.1}$~\cite{1307} & $ \chi_2(3 ^3 P_2)$, $\chi_2( 2 ^3 F_2)$, $\eta^h_{c2}( 1 ^1 D_2)$ \\ \hline
   $\eta_{c2}(1 D)$ & $2 ^{-+}$ & - & $ \eta_{c2}(1 ^1 D_2)$ \\ \hline
 \end{tabular}
\end{center}
\end{table}

\begin{table}\caption{Masses (in $\textrm{GeV}$) of hybrid charmonium mesons calculated by others with different models along with our calculated results. Our results are reported for least $J^{PC}$ states.}
\begin{center}
\begin{tabular}{|c|c|c|c|c|c|c|c|c|c|}
\hline
$J^{PC}$ state &Our results & Flux Tube & Constitute  &  \multicolumn{3}{|c|}{Lattice QCD} \\
& &Model &  glue model &  \multicolumn{2}{|c|}{ground state } & radial excited state \\
& &\cite{Isgur1} &\cite{IddirY4260} & \cite{Liu12}& \cite{12045425}&\cite{12045425}\\ \hline
$0^{--}$ &4.0802 & - & 4.35 &  - & - & \\
$0^{+-}$ & 4.22996 & 4.19 & - & $ \sim4.35$ & $\sim4.382$&- \\
$0^{-+}$ & 4.22996 & 4.19 & 4.38 & $ \sim4.120$ &-&- \\
$1^{-+}$ & 4.1063 & 4.19 & 4.48 & $ \sim4.300$ &$\sim 4.213$&\\
$1^{++}$ & 4.2678 & 4.19 & - &  - &-&-\\
$1^{+-}$ & 4.1063 & 4.19 & - & - &-&-\\
$1^{--}$ & 4.2678 & 4.19 & 4.27 & 4.189(54) &-&- \\
$2^{-+}$ & 4.2738 & 4.19 & - & $ \sim 4.300$ &-&-\\
$2^{+-}$ & 4.2738 & 4.19 & - & 4.4 &$\sim4.391$&$\sim 4.505$\\
$3^{-+}$ & 4.4197& -  & - & 4.7 & & \\ \hline
\end{tabular}
\end{center}
\end{table}

\begin{table}\caption{Root mean square radii and square of radial wave function at origin for ground and radially excited state
conventional charmonium mesons}
\begin{center}
\begin{tabular}{|c|c|c|c|c|c|c|c|c|c|}
\hline
n & Meson & $L$ & $S$ & J& $J^{PC}$ & our calculated & Theor. $\sqrt{\langle r^{2} \rangle}$ & Our calculated \\
& & & & &  & $\sqrt{ \langle r^{2} \rangle}$ &  with& $|R(0)|^2$ \\
& & & &  & &   & potential model~\cite{size}& \\ \hline
& & & &  &  & $ fm $ & $fm$ &$GeV^3$\\ \hline
1S & $ \eta_{c} (1 ^1S_0)$ & 0 & 0 & 0 &$0^{-+}$ &  0.3655 & 0.388& 1.2294 \\
 &$J/\psi (1 ^3S_1)$& $0$ & $1$ &1 &$1^{--}$ &$0.414$ & $0.404$& 1.97675\\ \hline
2 S & $ \eta_{c} (2 ^1S_0)$ & 0 & 0 & 0&$0^{-+}$ & 0.833 & ---&0.8717\\
 & $J/\psi (2 ^3S_1)$& $0$ & $1$ & 1 &$ 1^{--}$ &$0.863$ & --- &0.7225\\ \hline
3S & $ \eta_{c} (3 ^1S_0)$ & 0 & 0  & 0&$0^{-+}$ & 1.2072 & ---&.683\\
 & $J/\psi (3 ^3S_1)$& $0$ & $1$ &1 &$1^{--}$ &1.2287 & --- &.6006\\ \hline
4 S & $ \eta_{c} (4 ^1S_0)$ & 0 & 0 &0 &0$^{-+}$ & 1.5306 & ---&.5994\\
 & $J/\psi (4 ^3S_1)$& $0$ & $1$ &1 &$1^{--}$ & $1.5478$ & --- & .5417\\ \hline
5 S& $ \eta_{c} (5 ^1S_0)$ & 0 & 0 &0 &0$^{-+}$ &1.8224 & ---&.5503\\
 & $J/\psi (5 ^3S_1)$& $0$ & $1$ &1 &$1^{--}$ &$1.8370$ & --- &.50538\\ \hline
6 S& $ \eta_{c} (6 ^1S_0)$ & 0 & 0 &0 &0$^{-+}$ & 2.0922 & ---&0.5172\\
 & $J/\psi (6 ^3S_1)$& $0$ & $1$ &1 &$1^{--}$ & $2.1049$ & --- &0.4801\\ \hline
1P & $h_{c} (1 ^1P_1) $&  1 & $0$ & 1 &$1^{+-}$ & 0.6737 & 0.602&$\approx 0$\\
 & $\chi_{0} (1 ^3P_0)$ & $1$ & $1$ & 0 & $0^{++}$ & $0.621$ &$0.606$& $\approx 0$\\
 & $\chi_{1} (1 ^3P_1)$ & $1$ & $1$ &1 &$1^{++}$ & $0.673$ &---&$\approx 0$\\
 & $\chi_{2} (1 ^3P_2)$ & $1$ & $1$ &2 &$2^{++}$ & $0.717$ &---&$\approx 0$ \\ \hline
2 P& $h_{c} (2 ^1P_1) $&  1 & $0$ &1 & $1^{+-}$& $1.0697$ &---& $\approx 0$\\
 & $\chi_{0} (2 ^3P_0)$ & $1$ &1&0 &$ 0^{++}$ &$1.0374$ & ---&$\approx 0$\\
 & $\chi_{1} (2 ^3P_1)$ & $1$ & $1$ &1 &$1^{++}$ & $1.0729$ & ---&$\approx 0$\\
 & $\chi_{2} (2 ^3P_2)$ & $1$ & $1$ & 2&$2^{++}$ & $1.1106$ & ---& $\approx 0$\\ \hline
3 P & $h_{c} (3 ^1P_1) $&  1 & $0$ & 1 &$1^{+-}$ &$1.4052$ & ---&$\approx 0$ \\
 & $\chi_{0} (3 ^3P_0)$ & $1$ &1 &0 & $0^{++}$ &$1.3814$ & ---& $\approx 0$\\
 & $\chi_{1} (3 ^3P_1)$ & $1$ & $1$ &1 &$1^{++}$ &$1.4096$ & ---& $\approx 0$\\
 & $\chi_{2} (3 ^3P_2)$ & $1$ & $1$ &2 &$2^{++}$ & $1.4441$ & ---&$\approx 0$\\ \hline
4 P& $h_{c} (4 ^1P_1) $&  1 & $0$ & 1 &$1^{+-}$ & $1.7054$ & ---&$\approx 0$\\
 & $\chi_{0} (4 ^3P_0)$ & $1$ & $1$ &0 &$0^{++}$ & $1.6863$ & ---&$\approx 0$\\
 & $\chi_{1} (4 ^3P_1)$ & $1$ & $1$ &1  &$1^{++}$ & $1.7102$ & ---& $\approx 0$\\
 & $\chi_{2} (4 ^3P_2)$ & $1$ & $1$ & 2&$ 2^{++}$& $1.7427$ & ---&$\approx 0$\\ \hline
5 P & $h_{c} (5 ^1P_1) $&  1 & $0$ & 1 &$1^{+-}$ &$1.9815$ & ---&$\approx 0$\\
 & $\chi_{0} (5 ^3P_0)$ & $1$ & $1$ &0 &$0^{++}$ & $1.9654$ & ---& $\approx 0$\\
 & $\chi_{1} (5 ^3P_1)$ & $1$ & $1$ &1  &$1^{++}$ &$1.9863$ & ---& $\approx 0$\\
 & $\chi_{2} (5 ^3P_2)$ & $1$ & $1$ & 2&$ 2^{++}$&$2.0174$ & ---&$\approx 0$\\ \hline
6 P & $h_{c} (6 ^1P_1) $&  1 & $0$ & 1 &$1^{+-}$ &$2.23966$ & ---&$\approx 0$\\
 & $\chi_{0} (6 ^3P_0)$ & $1$ & $1$ &0 &$0^{++}$ & $2.2257$ & ---&$\approx 0$\\
 & $\chi_{1} (6 ^3P_1)$ & $1$ & $1$ &1  &$1^{++}$ & $2.2444$ & ---& $\approx 0$\\
 & $\chi_{2} (6 ^3P_2)$ & $1$ & $1$ & 2&$ 2^{++}$& $2.2745$ & ---&$\approx 0$\\ \hline
1 D & $\eta_{c2} (1 ^1D_2)$ & 2 & 0 & 2 &$2^{-+}$ &0.8984 &-&$\approx 0$\\
 &$\psi (1 ^3D_1)$& 2 & 1 &1 &$1^{--}$ & 0.8515 & $-$&$\approx 0$\\
 &$\psi_{2} (1 ^3D_2)$& 2 & 1 & 2 &$2^{--}$ &0.8937 &---&$\approx 0$\\
 &$\psi_{3} (1 ^3D_3)$& 2 & 1 &3 &$3^{--}$ &0.9182 & $---$& $\approx 0$\\ \hline
\end{tabular}
\end{center}
\end{table}
\begin{table}
\begin{center}
\begin{tabular}{|c|c|c|c|c|c|c|c|c|c|}
\hline
n & Meson & $L$ & $S$ & J& $J^{PC}$ & our calculated & Theor. $\sqrt{\langle r^{2} \rangle}$ & $|R(0)|^2$ \\
& & & & &  & $\sqrt{ \langle r^{2} \rangle}$ &  with& \\
& & & &  & &   & potential model~\cite{size}& \\ \hline
& & & &  &  & $fm$ & $fm$ &$GeV^3$\\ \hline
2 D & $ \eta_{c2}(2 ^1D_2)$ & 2 & 0 & 2 &$2^{-+}$ &1.2595 &---&$\approx 0$\\
 &$\psi (2 ^3D_1)$& 2 & 1 &1 &$1^{--}$ &  1.2112 & ---&$\approx 0$\\
 &$\psi_{2} (2 ^3D_2)$& 2 & 1 & 2 &$2^{--}$ & 1.2556 & ---&$\approx 0$\\
 &$\psi_{3} (2 ^3D_3)$& 2 & 1 &3 &$3^{--}$ & 1.2793 & --- & $\approx 0$ \\ \hline
3 D & $ \eta_{c2}(3 ^1D_2)$ & 2 & 0 & 2 &$2^{-+}$ & 1.57397 &--- & $\approx 0$\\
 &$\psi (3 ^3D_1)$& 2 & 1 &1 &$1^{--}$ &  1.52397 &--- & $\approx 0$ \\
 &$\psi_{2} (3 ^3D_2)$& 2 & 1 & 2 &$2^{--}$ &  1.5706 &--- & $\approx 0$\\
 &$\psi_{3} (3 ^3D_3)$& 2 & 1 &3 &$3^{--}$ & 1.59398 &--- & $\approx 0$\\ \hline
4 D & $ \eta_{c2}(4 ^1D_2)$ & 2 & 0 & 2 &$2^{-+}$ & 1.8596 & --- &$\approx 0$\\
 &$\psi (4 ^3D_1)$& 2 & 1 &1 &$1^{--}$ & 1.80785 & --- &$\approx 0$\\
 &$\psi_{2} (4 ^3D_2)$& 2 & 1 & 2 &$2^{--}$ & 1.8565 &--- & $\approx 0$\\
 &$\psi_{3} (4 ^3D_3)$& 2 & 1 &3 &$3^{--}$ & 1.8798 &--- &$\approx 0$\\ \hline
5 D & $ \eta_{c2}(5 ^1D_2)$ & 2 & 0 & 2 &$2^{-+}$ & 2.1248 &--- & $\approx 0$\\
 &$\psi (5 ^3D_1)$& 2 & 1 &1 &$1^{--}$ & 2.0714 & --- &$\approx 0$\\
 &$\psi_{2} (5 ^3D_2)$& 2 & 1 & 2 &$2^{--}$ & 2.1218 & --- &$\approx 0$\\
 &$\psi_{3} (5 ^3D_3)$& 2 & 1 &3 &$3^{--}$ & 2.1452 & --- &$\approx 0$\\ \hline
1 F& $h_{c3}(1 ^1F_3) $&  3 & $0$ & 3 &$3^{+-}$ & 1.0878 & ---&$\approx 0$ \\
 & $\chi_2 (1 ^3F_2)$ & $3$ & $1$ &2 &$2^{++}$ & 1.06896 & ---&$\approx 0$\\
 & $\chi_3(1 ^3F_3)$ & $3$ & $1$ &3  &$3^{++}$ & $1.08628$ & ---& $\approx 0$\\
 & $\chi_4(1 ^3F_4)$ & $3$ & $1$ & 4 &$ 4^{++}$&1.09818 & ---& $\approx 0$\\ \hline
2 F& $h_{c3}(2 ^1F_3) $&  3 & $0$ & 3 &$3^{+-}$ & $1.4253$ & ---&$\approx 0$ \\
 & $\chi_2(2 ^3F_2)$ & $3$ & $1$ &2 &$2^{++}$ & 1.4062 & ---&$\approx 0$ \\
 & $\chi_3(2 ^3F_3)$ & $3$ & $1$ &3  &$3^{++}$ & $1.42387$ & ---&$\approx 0$\\
 & $\chi_4(2 ^3F_4)$ & $3$ & $1$ & 4 &$ 4^{++}$& 1.43578 & ---& $\approx 0$\\ \hline
3 F& $h_{c3}(3 ^1F_3) $&  3 & $0$ & 3 &$3^{+-}$ &$1.725$ & ---&$\approx 0$ \\
 & $\chi_2(3 ^3F_2)$ & $3$ & $1$ &2 &$2^{++}$ & $1.7056$ & ---&$\approx 0$\\
 & $\chi_3(3 ^3F_3)$ & $3$ & $1$ &3  &$3^{++}$ & $1.72369$ & ---&$\approx 0$ \\
 & $\chi_4(3 ^3F_4)$ & $3$ & $1$ & 4 &$ 4^{++}$& $1.73579$ & ---&$\approx 0$\\ \hline
4 F& $h_{c3}(4 ^1F_3) $&  3 & $0$ & 3 &$3^{+-}$ & $2.0$ & ---&$\approx 0$\\
 & $\chi_2(4^3F_2)$ & $3$ & $1$ &2 &$2^{++}$ &$1.9802$ & --- &$\approx 0$\\
 & $\chi_3(4 ^3F_3)$ & $3$ & $1$ &3  &$3^{++}$ &$1.9988$ & --- &$\approx 0$\\
 & $\chi_4(4 ^3F_4)$ & $3$ & $1$ & 4 &$ 4^{++}$&$2.0111$ & ---&$\approx 0$\\ \hline
1 G& $ \eta_{c4}(1 ^1G_4)$ & 4 & 0 & 4 &$4^{-+}$ & 1.2589 & --- &$\approx 0$\\
 &$\psi_3 (1 ^3G_3)$& 4 & 1 &3 &$3^{--}$ & 1.25026 & --- &$\approx 0$\\
 &$\psi_4 (1 ^3G_4)$& 4 & 1 & 4 &$4^{--}$ & 1.25878 & --- &$\approx 0$\\
 &$\psi_5 (1 ^3G_5)$& 4 & 1 &5 &$5^{--}$ & 1.26409 & --- &$\approx 0$\\ \hline
\end{tabular}
\end{center}
\end{table}

\begin{table}\caption{Our calculated
root mean square radii and $|R(0)|^2$ of $c\overline{c}$ hybrid mesons.}
\begin{center}
\begin{tabular}{|c|c|c|c|c|c|c|c|c|}
\hline
n & Meson & $L$ & $S$ & J& \multicolumn{2}{|c|}{$J^{PC}$} & our calculated & Our calculated   \\
& & & & &$\varepsilon=1$ &$\varepsilon=-1$ & $\sqrt{ \langle r^{2} \rangle}$ & $|R(0)|^2$  \\
& & & &  & & & &  \\ \hline
& & & &  & & & $fm$ & $Gev^3$\\ \hline
1S & $ \eta^{h}_{c} (1 ^1S_0)$ & 0 & 0 & 0 &$0^{++}$ & $0^{--}$ & 0.6429& .30458\\
 &$ J/\psi^{h} (1 ^3S_1) $& $0$ & $1$ &1 &$1^{+-}$ & $1^{-+}$ &0.6949 & 0.1533\\ \hline
2 S & $ \eta'^{h}_{c} (2 ^1S_0)$ & 0 & 0 & 0&$0^{++}$ & $0^{--}$ & 1.0837& 0.3306\\
 & $ J/\psi^{h} (2 ^3S_1)$& $0$ & $1$ & 1 &$ 1^{+-}$ &  $1^{-+}$ &1.1187& 0.1995 \\ \hline
3S & $ \eta^{h}_{c} (3 ^1S_0)$ & 0 & 0  & 0&$0^{++}$ & $0^{--}$ & 1.4345& 0.3259 \\
 & $J/\psi^{h} (3 ^3S_1)$& $0$ & $1$ &1 &$1^{+-}$ &  $1^{-+}$ &1.4609 & 0.2214\\ \hline
4 S & $ \eta^{h}_{c} (4 ^1S_0)$ & 0 & 0 &0 &$0^{++}$ & $0^{--}$ & 1.7413 &0.3189\\
 & $J/\psi^{h} (4 ^3S_1)$& $0$ & $1$ &1 &$1^{+-}$ &  $1^{-+}$ &$1.7624$ &0.2342 \\ \hline
5 S& $ \eta^{h}_{c} (5 ^1S_0)$ & 0 & 0 & 0 &$0^{++}$ & $0^{--}$ & 2.0203&0.3128\\
 & $J/\psi^{h} (5 ^3S_1)$& $0$ & $1$ &1 &$1^{+-}$ &  $1^{-+}$ &2.0379 &0.2425\\ \hline
6 S& $ \eta^{h}_{c} (6 ^1S_0)$ & 0 & 0 &0 &$0^{++}$ & $0^{--}$ & 2.2797&0.30778 \\
 & $J/\psi^{h} (6 ^3S_1)$& $0$ & $1$ &1 &$1^{+-}$ &  $1^{-+}$ & 2.2948 &0.2482 \\ \hline
1P & $ h^{h}_{c} (1 ^1P_1) $&  1 & $0$ & 1 &$1^{--}$ & $1^{++}$ & .922397&$\approx 0$ \\
 & $\chi^{h}_{0} (1 ^3P_0)$ & $1$ & $1$ & 0 & $0^{-+}$ & $0^{+-}$ &0.8765&$\approx 0$\\
 & $\chi^{h}_{1} (1 ^3P_1)$ & $1$ & $1$ &1 &$1^{-+}$ &$1^{+-}$ & 0.9117&$\approx 0$\\
 & $\chi^{h}_{2} (1 ^3P_2)$ & $1$ & $1$ &2 &$2^{-+}$&$ 2^{+-}$&0.9453 &$\approx 0$\\ \hline
2 P& $ h^{h}_{c} (2 ^1P_1) $&  1 & $0$ &1 & $1^{--}$ & $1^{++}$ & 1.2964&$\approx 0$\\
 & $\chi^{h}_{0} (2 ^3P_0)$ & $1$ &1&0 &$ 0^{-+}$ &$ 0^{+-}$ &1.2616 &$\approx 0$\\
 & $\chi^{h}_{1} (2 ^3P_1)$ & $1$ & $1$ &1 &$1^{-+}$ &$1^{+-}$ & 1.2848&$\approx 0$\\
 & $\chi^{h}_{2} (2 ^3P_2)$ & $1$ & $1$ & 2&$2^{-+}$ &$2^{-+}$& 1.3210&$\approx 0$\\ \hline
3 P & $h^{h}_{c} (3 ^1P_1) $&  1 & $0$ & 1 &$1^{--}$  & $1^{++}$ &1.6139&$\approx 0$\\
 & $\chi^{h}_{0} (3 ^3P_0)$ & $1$ &1 &0 & $0^{-+}$ &$ 0^{+-}$ & 1.5875& $\approx 0$\\
 & $\chi^{h}_{1} (3 ^3P_1)$ & $1$ & $1$ &1 &$1^{-+}$ &$1^{+-}$ & 1.6032& $\approx 0$\\
 & $\chi^{h}_{2} (3 ^3P_2)$ & $1$ & $1$ &2 &$2^{-+}$ &$2^{-+}$& 1.6410 & $\approx 0$\\ \hline
4 P& $h^{h}_{c} (4 ^1P_1) $&  1 & $0$ & 1 &$1^{--}$ & $1^{++}$ & 1.9009& $\approx 0$\\
 & $\chi^{h}_{0} (4 ^3P_0)$ & $1$ & $1$ &0 &$0^{-+}$ &$ 0^{+-}$ &1.8793& $\approx 0$\\
 & $\chi^{h}_{1} (4 ^3P_1)$ & $1$ & $1$ &1  &$1^{-+}$ &$1^{+-}$ & 1.8898& $\approx 0$\\
 & $\chi^{h}_{2} (4 ^3P_2)$ & $1$ & $1$ & 2&$ 2^{-+}$&$2^{-+}$&1.9287 & $\approx 0$\\ \hline
5 P & $h^{h}_{c} (5 ^1P_1) $&  1 & $0$ & 1 &$1^{1--}$  & $1^{++}$ & 2.1663 & $\approx 0$\\
 & $\chi^{h}_{0} (5 ^3P_0)$ & $1$ & $1$ &0 &$0^{-+}$ &$ 0^{+-}$ & 2.1481 & $\approx 0$\\
 & $\chi^{h}_{1} (5 ^3P_1)$ & $1$ & $1$ &1  &$1^{-+}$ &$1^{+-}$ & 2.1548 & $\approx 0$\\
 & $\chi^{h}_{2} (5 ^3P_2)$ & $1$ & $1$ & 2&$ 2^{-+}$&$2^{-+}$& $2.1944$ & $\approx 0$\\ \hline
6 P & $h^{h}_{c} (6 ^1P_1) $&  1 & $0$ & 1 &$1^{--}$ & $1^{++}$ & 2.4155 & $\approx 0$\\
 & $\chi^{h}_{0} (6 ^3P_0)$ & $1$ & $1$ &0 &$0^{-+}$ &$ 0^{+-}$ & $2.3998$& $\approx 0$\\
 & $\chi^{h}_{1} (6 ^3P_1)$ & $1$ & $1$ &1  &$1^{-+}$ &$1^{+-}$ & $2.4022$ & $\approx 0$\\
 & $\chi^{h}_{2} (6 ^3P_2)$ & $1$ & $1$ & 2&$ 2^{-+}$&$2^{-+}$& $2.4437$ & $\approx 0$\\ \hline
1 D & $ \eta^{h}_{c2}(1 ^1D_2)$ & 2 & 0 & 2 &$2^{++}$ &$2^{--}$& 1.1156 & $\approx 0$\\
 &$\psi^{h} (1 ^3D_1)$& 2 & 1 &1 &$1^{+-}$ & $1^{-+}$ & 1.0923 & $\approx 0$\\
 &$\psi^{h}_{2} (1 ^3D_2)$& 2 & 1 & 2 &$2^{+-}$ & $2^{-+}$ &1.1133 & $\approx 0$\\
 &$\psi^{h}_{3} (1 ^3D_3)$& 2 & 1 &3 &$3^{+-}$ & $3^{-+}$ & 1.1263 & $\approx 0$ \\ \hline
2 D & $ \eta^{h}_{c2}(2 ^1D_2)$ & 2 & 0 & 2 &$2^{++}$ &$2^{--}$&1.4613 & $\approx 0$\\
 &$\psi^{h} (2 ^3D_1)$& 2 & 1 &1 &$1^{+-}$ &$1^{-+}$ &1.4359 & $\approx 0$\\
 &$\psi^{h}_{2} (2 ^3D_2)$& 2 & 1 & 2 &$2^{+-}$ & $2^{-+}$ &1.4589& $\approx 0$\\
 &$\psi^{h}_{3} (2 ^3D_3)$& 2 & 1 &3 &$3^{+-}$ &  $3^{-+}$ &1.4733& $\approx 0$\\ \hline
\end{tabular}
\end{center}
\end{table}
\begin{table}
\begin{center}
\begin{tabular}{|c|c|c|c|c|c|c|c|c|}
\hline
n & Meson & $L$ & $S$ & J& \multicolumn{2}{|c|}{$J^{PC}$} & our calculated &our calculated \\
& & & & &$\varepsilon=1$ & $\varepsilon=-1$& $\sqrt{ \langle r^{2} \rangle}$ &$|R(0)|^2$ \\
& & & & & &  & &\\ \hline
& & & & & &  & $fm$& $GeV^3$ \\ \hline
3 D & $ \eta^{h}_{c2}(3 ^1D_2)$ & 2 & 0 & 2 &$2^{++}$ &$2^{--}$& 1.7642& $\approx 0$\\
 &$\psi^{h} (3 ^3D_1)$& 2 & 1 &1 &$1^{+-}$ & $1^{-+}$ &1.7376 &$\approx 0$\\
 &$\psi^{h}_{2} (3 ^3D_2)$& 2 & 1 & 2 &$2^{+-}$ &$2^{-+}$ & 1.7618& $\approx 0$\\
 &$\psi^{h}_{3} (3 ^3D_3)$& 2 & 1 &3 &$3^{+-}$ & $3^{-+}$ &1.7773 & $\approx 0$\\ \hline
4 D & $ \eta^{h}_{c2}(4 ^1D_2)$ & 2 & 0 & 2 &$2^{++}$ &$2^{--}$&2.0403&$\approx 0$\\
 &$\psi^{h} (4 ^3D_1)$& 2 & 1 &1 &$1^{+-}$ &$1^{-+}$ &2.01288& $\approx 0$ \\
 &$\psi^{h}_{2} (4 ^3D_2)$& 2 & 1 & 2 &$2^{+-}$ &$2^{-+}$ &2.03804 & $\approx 0$\\
 &$\psi^{h}_{3} (4 ^3D_3)$& 2 & 1 &3 &$3^{+-}$ & $3^{-+}$ &2.0543& $\approx 0$ \\ \hline
5 D & $ \eta^{h}_{c2}(5 ^1D_2)$ & 2 & 0 & 2 &$2^{++}$ &$2^{--}$& 2.2974& $\approx 0$\\
 &$\psi^{h} (5 ^3D_1)$& 2 & 1 &1 &$1^{+-}$ &$1^{-+}$ & 2.2695& $\approx 0$\\
 &$\psi^{h}_{2} (5 ^3D_2)$& 2 & 1 & 2 &$2^{+-}$ & $2^{-+}$ &2.2952& $\approx 0$\\
 &$\psi^{h}_{3} (5 ^3D_3)$& 2 & 1 &3 &$3^{+-}$ & $3^{-+}$ &2.3121& $\approx 0$\\ \hline
1 F & $h^{h}_{c3}(1 ^1 F_3) $&  3 & $0$ & 3 &$3^{--}$ & $3^{++}$ &1.2857& $\approx 0$\\
 & $\chi^{h}_{2}(1 ^3 F_2)$ & $3$ & $1$ &2 &$2^{-+}$ &$2^{+-}$ &1.2759&$\approx 0$\\
 & $\chi^{h}_{3}(1 ^3 F_3)$ & $3$ & $1$ &3 &$3^{-+}$ & $3^{+-}$ &1.2855& $\approx 0$\\
 & $\chi^{h}_{4}(1 ^3 F_4)$ & $3$ & $1$ & 4&$ 4^{-+}$& $ 4^{+-}$&1.2908& $\approx 0$\\ \hline
2 F & $h^{h}_{c3}(2 ^1 F_3) $&  3 & $0$ & 3 &$3^{--}$ &$3^{++}$ & 1.611& $\approx 0$\\
 & $\chi^{h}_{2}(2 ^3 F_2)$ & $3$ & $1$ &2 &$2^{-+}$ &$2^{+-}$ &1.5997 & $\approx 0$\\
 & $\chi^{h}_{3}(2 ^3 F_3)$ & $3$ & $1$ &3 &$3^{-+}$ &$3^{+-}$ & 1.6105& $\approx 0$\\
 & $\chi^{h}_{4}(2 ^3 F_4)$ & $3$ & $1$ & 4&$ 4^{-+}$&$ 4^{+-}$& 1.6171& $\approx 0$\\ \hline
3 F & $h^{h}_{c3}(3 ^1 F_3) $&  3 & $0$ & 3 &$3^{--}$ &$3^{++}$ & 1.9015& $\approx 0$\\
 & $\chi^{h}_{2}(3 ^3 F_2)$ & $3$ & $1$ &2 &$2^{-+}$ &$2^{+-}$ &1.88895 & $\approx 0$\\
 & $\chi^{h}_{3}(3 ^3 F_3)$ & $3$ & $1$ &3 &$3^{-+}$ &$3^{+-}$ &1.90084& $\approx 0$\\
 & $\chi^{h}_{4}(3 ^3 F_4)$ & $3$ & $1$ & 4&$ 4^{-+}$& $ 4^{+-}$ & 1.9083& $\approx 0$\\ \hline
4 F & $h^{h}_{c3}(4 ^1 F_3) $&  3 & $0$ & 3 &$3^{--}$ &$3^{++}$ &2.1688& $\approx 0$\\
 & $\chi^{h}_{2}(4 ^3 F_2)$ & $3$ & $1$ &2 &$2^{-+}$ &$2^{+-}$ &2.1553& $\approx 0$\\
 & $\chi^{h}_{3}(4 ^3 F_3)$ & $3$ & $1$ &3 &$3^{-+}$ &$3^{+-}$ &2.1681& $\approx 0$\\
 & $\chi^{h}_{4}(4 ^3 F_4)$ & $3$ & $1$ & 4&$ 4^{-+}$& $ 4^{+-}$& 2.1762& $\approx 0$\\ \hline
\end{tabular}
\end{center}
\end{table}
\begin{table}
\begin{center}
\begin{tabular}{|c|c|c|c|c|c|c|c|c|}
\hline
n & Meson & $L$ & $S$ & J& \multicolumn{2}{|c|}{$J^{PC}$} & our calculated & Our calculated   \\
& & & & &$\varepsilon=1$ &$\varepsilon=-1$ & $\sqrt{ \langle r^{2} \rangle}$ & $|R(0)|^2$  \\
& & & &  & & & &  \\ \hline
& & & &  & & & $GeV^{-1}$ & $Gev^3$\\ \hline
4 D & $ \eta_{c2}(4 ^1D_2)$ & 2 & 0 & 2 &$2^{-+}$ & 1.8596 &--- &$\approx 0$ \\
 &$\psi (4 ^3D_1)$& 2 & 1 &1 &$1^{--}$ & 1.80785 &--- &$\approx 0$\\
 &$\psi_{2} (4 ^3D_2)$& 2 & 1 & 2 &$2^{--}$ & 1.8565 & --- &$\approx 0$\\
 &$\psi_{3} (4 ^3D_3)$& 2 & 1 &3 &$3^{--}$ & 1.8798 & --- &$\approx 0$\\ \hline
5 D & $ \eta_{c2}(5 ^1D_2)$ & 2 & 0 & 2 &$2^{-+}$ & 2.1248 &--- &$\approx 0$\\
 &$\psi (5 ^3D_1)$& 2 & 1 &1 &$1^{--}$ & 2.0714 & --- &$\approx 0$ \\
 &$\psi_{2} (5 ^3D_2)$& 2 & 1 & 2 &$2^{--}$ & 2.1218 & --- &$\approx 0$\\
 &$\psi_{3} (5 ^3D_3)$& 2 & 1 &3 &$3^{--}$ & 2.1452 & --- &$\approx 0$\\ \hline
1 F& $h_{c3}(1 ^1F_3) $&  3 & $0$ & 3 &$3^{+-}$ & 1.0878 &--- &$\approx 0$ \\
 & $\chi_2 (1 ^3F_2)$ & $3$ & $1$ &2 &$2^{++}$ & 1.06896 & --- &$\approx 0$ \\
 & $\chi_3(1 ^3F_3)$ & $3$ & $1$ &3  &$3^{++}$ & $1.08628$ &--- &$\approx 0$ \\
 & $\chi_4(1 ^3F_4)$ & $3$ & $1$ & 4 &$ 4^{++}$&1.09818 &--- &$\approx 0$\\ \hline
2 F& $h_{c3}(2 ^1F_3) $&  3 & $0$ & 3 &$3^{+-}$ & $1.4253$ & --- &$\approx 0$ \\
 & $\chi_2(2 ^3F_2)$ & $3$ & $1$ &2 &$2^{++}$ & 1.4062 &--- &$\approx 0$\\
 & $\chi_3(2 ^3F_3)$ & $3$ & $1$ &3  &$3^{++}$ & $1.42387$ &--- &$\approx 0$\\
 & $\chi_4(2 ^3F_4)$ & $3$ & $1$ & 4 &$ 4^{++}$& 1.43578 & --- &$\approx 0$\\ \hline
3 F& $h_{c3}(3 ^1F_3) $&  3 & $0$ & 3 &$3^{+-}$ &$1.725$ & --- &$\approx 0$\\
 & $\chi_2(3 ^3F_2)$ & $3$ & $1$ &2 &$2^{++}$ & $1.7056$ & --- &$\approx 0$\\
 & $\chi_3(3 ^3F_3)$ & $3$ & $1$ &3  &$3^{++}$ & $1.72369$ & --- &$\approx 0$ \\
 & $\chi_4(3 ^3F_4)$ & $3$ & $1$ & 4 &$ 4^{++}$& $1.73579$ & --- &$\approx 0$\\ \hline
4 F& $h_{c3}(4 ^1F_3) $&  3 & $0$ & 3 &$3^{+-}$ & $2.00004$ &--- &$\approx 0$\\
 & $\chi_2(4^3F_2)$ & $3$ & $1$ &2 &$2^{++}$ &$1.9802$ & --- &$\approx 0$\\
 & $\chi_3(4 ^3F_3)$ & $3$ & $1$ &3  &$3^{++}$ &$1.9988$ &--- &$\approx 0$\\
 & $\chi_4(4 ^3F_4)$ & $3$ & $1$ & 4 &$ 4^{++}$&$2.0111$ &--- &$\approx 0$\\ \hline
1 G& $ \eta_{c4}(1 ^1G_4)$ & 4 & 0 & 4 &$4^{-+}$ & 1.2589 &--- &$\approx 0$\\
 &$\psi_3 (1 ^3G_3)$& 4 & 1 &3 &$3^{--}$ & 1.25026 & --- &$\approx 0$\\
 &$\psi_4 (1 ^3G_4)$& 4 & 1 & 4 &$4^{--}$ & 1.25878 & --- &$\approx 0$\\
 &$\psi_5 (1 ^3G_5)$& 4 & 1 &5 &$5^{--}$ & 1.26409 & --- &$\approx 0$\\ \hline
\end{tabular}
\end{center}
\end{table}

\section*{Acknowledgement}
\qquad B. M. and F. A. acknowledge the financial support of Punjab University for the projects(Sr. 215 PU Project 2012-13 and Sr. 220 PU Project 2012-13). N. A. and B. M. are grateful to Higher education Commission of Pakistan for their financial support.

\end{document}